\documentclass{llncs}

\usepackage{url,amsmath,amssymb}

\usepackage{stmaryrd}

\newbox\tempa
\newbox\tempb
\newdimen\tempc
\def\mud#1{\hfil $\displaystyle{\mathstrut #1}$\hfil}
\def\rig#1{\hfil $\displaystyle{#1}$}
\def\irulehelp#1#2#3{\setbox\tempa=\hbox{$\displaystyle{\mathstrut #2}$}%
                        \setbox\tempb=\vbox{\halign{##\cr
        \mud{#1}\cr
        \noalign{\vskip\the\lineskip}%
        \noalign{\hrule height 0pt}%
        \rig{\vbox to 0pt{\vss\hbox to 0pt{${\; #3}$\hss}\vss}}\cr
        \noalign{\hrule}%
        \noalign{\vskip\the\lineskip}%

        \mud{\copy\tempa}\cr}}%
                      \tempc=\wd\tempb
                      \advance\tempc by \wd\tempa
                      \divide\tempc by 2 }
\def\irule#1#2#3{{\irulehelp{#1}{#2}{#3}%
                     \hbox to \wd\tempa{\hss \box\tempb \hss}}}

\newcommand{\lra}{\longrightarrow}
\newcommand{\fa}{\forall}
\newcommand{\ex}{\exists}

\title{Deduction modulo theory} 

\author{Gilles Dowek}

\authorrunning{Gilles Dowek}

\institute{Inria, 23 avenue d'Italie, CS 81321, 75214 Paris Cedex 13, France.\\
 \email{gilles.dowek@inria.fr}}

\begin{document}

\maketitle

\section{Introduction}

\subsection{Weaker vs. stronger systems}

Contemporary proof theory goes into several directions at the same time.
One of them aims at analyzing proofs,
propositions, connectives, etc., that is at decomposing them
into more atomic objects.  This often leads to design systems that are
weaker than Predicate logic, but that have better algebraic or
computational properties, and to try to reconstruct part of Predicate
logic on top of these systems.  Propositional logic, linear logic,
deep inference, equational logic, explicit substitution calculi,
etc. are examples of such systems.  From this point of view, Predicate
logic appears more as the ultimate goal of the journey, than as its
starting point.

Another direction considers that very little can be expressed in pure
Predicate logic and that stronger systems are needed, for instance to
express genuine mathematical proofs. Axiomatic theories, modal logics,
type theories, etc. are examples of such systems that are more
expressive than pure Predicate logic.  
There, Predicate logic is the starting point of the journey.

Although both points of view coexist in many research projects, these
two approaches to proof theory often lead to different systems
and different problems.

{\em Deduction modulo theory} is part of the second group, as it
focuses on proofs in theories. The concern of integrating theories
to proof theory is that of several research groups. See, for
instance, \cite{NegriPlato} and \cite{Naibo} for related approaches.

\subsection{Logical vs. theoretical systems}

To design a system stronger than pure Predicate logic, several ways
are possible. One is to extend Predicate logic with new logical
operators, that is to design a logic, the second is to 
introduce function symbols and predicate symbols
within Predicate logic and state axioms expressing the meaning of these
symbols, that is to design a theory. The first approach can be
illustrated by modal logics, the second by arithmetic or set
theory. Simple type theory belongs to both groups as it
can be defined either as a logic, in which
case it is more often called {\em higher-order logic}, or as a theory in
Predicate logic \cite{DowekSkolemization}.

Deduction modulo theory is part of the second, theoretical rather than
logical, group, as, like Predicate logic, it is a framework in which 
it is possible to define many theories.

\subsection{Axioms vs. reduction rules}

But, the main difference between Deduction modulo theory and the axiomatic
approach is that a, in Deduction modulo theory, {\em theory} 
is not defined as a set of axioms, but as a set of reduction rules.

Indeed, axioms jeopardize most of the properties of proofs of pure
Predicate logic.  For instance, in pure Predicate logic, a constructive
Natural deduction 
cut free proof always ends with an introduction rule, hence a
constructive cut free existential proof always ends with an introduction rule
of the existential quantifier. But this result does not extend to
axiomatic theories, as a constructive cut free proof in a theory may also
end, for instance, with the axiom rule.

In the same way, in automated theorem proving in pure Predicate logic, 
the search space of the proposition $\bot$ is always finite. But this result
does not extend to axiomatic theories, that can generate an infinite search 
space for the proposition $\bot$.

To overcome these problems, axioms, 
in Deduction modulo theory, are replaced by sets of reduction rules. 
For instance, the axioms 
$$\fa y~(0 + y = y)$$
$$\fa x \fa y~(S(x) + y = S(x + y))$$
are replaced by the reduction rules 
$$0 + y \lra y$$
$$S(x) + y \lra S(x + y)$$
These reduction rules define a congruence $\equiv$ on propositions,
and deduction is performed modulo this congruence.
For instance, with the reduction rules above the propositions 
$2 + 2 = 4$ and $4 = 4$ are congruent, hence any proof of the latter is a 
proof of the former.
If, to define equality, we add the following rules \cite{Allali},
which directly rewrite atomic propositions
$$0 = 0 \lra \top$$
$$S(x) = 0 \lra \bot$$
$$0 = S(y) \lra \bot$$
$$S(x) = S(y) \lra x = y$$
then the proposition 
$2 + 2 = 4$ and $\top$ are congruent, and any proof of $\top$---for 
instance the mere application of the introduction rule for $\top$---is 
a proof of the proposition $2 + 2 = 4$
$$\irule{}
        {\vdash 2 + 2 = 4}
        {\mbox{$\top$-intro}}$$

\subsection{Deduction vs. computation}

In the example above, the proposition $2 + 2 = 4$ is provable because
it reduces to $\top$. More generally, all propositions that reduce to 
$\top$ are provable. But the converse is not true: not all provable 
propositions reduce to $\top$.  Indeed, reducibility to $\top$ is often a 
decidable property,
while provability is not.

On the contrary, the fact that the proposition $2 + 2 = 4$ has a
trivial proof, because it reduces to $\top$, 
shows that the truth of this proposition rests
on a mere computation and not on a genuine deduction. 

Thus, Deduction modulo theory also permits one to distinguish
the computation part from the deduction part
within a proof, whereas
Predicate logic
flattens computation and deduction at the same level.

\subsection{The origins of Deduction modulo theory}

Deduction modulo theory was first introduced in the area of 
automated theorem proving. 

Indeed, in Resolution, or in other automated theorem proving methods, 
instead of using equational
axioms, for instance the associativity axiom, we often replace standard
unification with equational unification, for instance unification
modulo associativity \cite{Plotkin}.  In the same
way, in Simple type theory, instead of using the $\beta$-conversion
axiom, we replace standard unification with equational
unification modulo $\beta$-equivalence:
higher-order unification \cite{Andrews71,Huet73,Huet75}. The 
automated theorem proving method obtained this way is called {\em Equational 
resolution}. 

A way to prove the soundness and completeness of Equational resolution
is to introduce a Natural deduction system, or a Sequent calculus
system, where propositions are identified modulo associativity, or
modulo $\beta$-equivalence. Then, this system can be proved to be
equivalent to the axiomatic presentation of the theory.  Finally, the
soundness and completeness of Equational resolution are proved
relatively to this system, where every deduction step is performed
modulo the congruence.

So Deduction modulo theory comes from automated theorem proving.  But
it was soon understood that this idea of identifying propositions
modulo a congruence was also the idea behind the notion of
definitional equality in Martin-L\"of's Intuitionistic type theory
\cite{MartinLof} and that Deduction modulo theory could also be seen
as an extension of this notion of definitional equality to Predicate
logic.

Another source of inspiration is the extension of 
Natural deduction with folding and unfolding rules, introduced by Prawitz
\cite{Prawitz,Crabbe74,Hallnas,Ekman,Crabbe91}.
In this system, it is not possible to identify an atomic proposition $P$
with a proposition $A$. But, it is possible to introduce non logical 
deduction rules 
$$\begin{array}{cc}
\irule{A}{P}{} ~~~~~~~~~~~~~~~~~& \irule{P}{A}{}
\end{array}$$ 
The relation between the two frameworks is detailed in 
\cite{foldunfold}.

\section{Proof Systems}

The idea of reasoning modulo a theory can be used in different
formalisms: Natural deduction, Sequent calculus, $\lambda$-calculus, etc.
Thus, Deduction modulo theory exists in many flavors. 

\subsection{Natural Deduction modulo theory}

Let us start with constructive Natural deduction.  The rules of
constructive Natural deduction modulo theory are obtained by
transforming the rules of constructive Natural deduction, to allow to
use of the congruence. For instance, the rule
$$\irule{\Gamma \vdash A \Rightarrow B~~~\Gamma \vdash A}
        {\Gamma \vdash B}
        {\mbox{$\Rightarrow$-elim}}$$
is transformed into 
$$\irule{\Gamma \vdash C~~~\Gamma \vdash A}
        {\Gamma \vdash B}
        {\begin{array}{l} \mbox{$\Rightarrow$-elim}\\ \mbox{if $C \equiv (A \Rightarrow B)$}\end{array}}$$
where the proposition $A \Rightarrow B$ is replaced by any congruent 
proposition $C$. 
Applying the same transformation to all Natural deduction rules yields
the system of Figure \ref{ndm}.

\begin{figure}
$$\hspace*{-4cm}
\begin{array}{cc}
\irule{}
      {\Gamma, A \vdash B}
      {\begin{array}{l} \mbox{axiom}\\ \mbox{if $A \equiv B$}\end{array}}
\\
\\

\irule{}
      {\Gamma \vdash A}
      {\begin{array}{l} \mbox{$\top$-intro}\\ \mbox{if $A \equiv \top$} \end{array}}
~~~~~~~~~~~~~~~~~~~~~~~~~~~~~~~~~~~~~~~~~~~~~~~~~~~~~
\\
\\

& 
\irule{\Gamma \vdash B}
      {\Gamma \vdash A}
      {\begin{array}{l} \mbox{$\bot$-elim}\\ \mbox{if $B \equiv \bot$} \end{array}}
\\
\\

\irule{\Gamma \vdash A~~~\Gamma \vdash B}
      {\Gamma \vdash C}
      {\begin{array}{l} \mbox{$\wedge$-intro}\\ \mbox{if $C \equiv (A \wedge B)$}\end{array}}
~~~~~~~~~~~~~~~~~~~~~~~~~~~~~~~~~~~~~~~~~~~~~~~~~~~~~& 
\irule{\Gamma \vdash C}
      {\Gamma \vdash A}
      {\begin{array}{l} \mbox{$\wedge$-elim}\\ \mbox{ if $C \equiv (A \wedge B)$}\end{array}}

\\
\\

~~~~~~~~~~~~~~~~~~~~~~~~~~~~~~~~~~~~~~~~~~~~~~~~~~~~~& 
\irule{\Gamma \vdash C}
      {\Gamma \vdash B}
      {\begin{array}{l} \mbox{$\wedge$-elim}\\ \mbox{if $C \equiv (A \wedge B)$}\end{array}}
\\
\\

\irule{\Gamma \vdash A}
      {\Gamma \vdash C}
      {\begin{array}{l} \mbox{$\vee$-intro}\\ \mbox{if $C \equiv (A \vee B)$}\end{array}}
~~~~~~~~~~~~~~~~~~~~~~~~~~~~~~~~~~~~~~~~~~~~~~~~~~~~~& 
\irule{\Gamma \vdash D~~~\Gamma, A \vdash C~~~\Gamma, B \vdash C}
      {\Gamma \vdash C}
      {\begin{array}{l} \mbox{$\vee$-elim}\\ \mbox{if $D \equiv (A \vee B)$}\end{array}}
\\
\\

\irule{\Gamma \vdash B}
      {\Gamma \vdash C}
      {\begin{array}{l} \mbox{$\vee$-intro}\\ \mbox{if $C \equiv (A \vee B)$}\end{array}}
~~~~~~~~~~~~~~~~~~~~~~~~~~~~~~~~~~~~~~~~~~~~~~~~~~~~~
\\
\\

\irule{\Gamma, A \vdash B}
      {\Gamma \vdash C}
      {\begin{array}{l} \mbox{$\Rightarrow$-intro}\\ \mbox{if $C \equiv (A \Rightarrow B)$}\end{array}}
~~~~~~~~~~~~~~~~~~~~~~~~~~~~~~~~~~~~~~~~~~~~~~~~~~~~~& 
\irule{\Gamma \vdash C~~~\Gamma \vdash A}
      {\Gamma \vdash B}
      {\begin{array}{l} \mbox{$\Rightarrow$-elim}\\ \mbox{if $C \equiv (A \Rightarrow B)$}\end{array}}
\\
\\

\irule{\Gamma \vdash A}
      {\Gamma \vdash B}
      {\begin{array}{l} \mbox{$\langle x,A \rangle$ $\fa$-intro}\\ \mbox{if $B \equiv (\fa x~A)$}\\ \mbox{and $x \not\in FV(\Gamma)$} \end{array}}
~~~~~~~~~~~~~~~~~~~~~~~~~~~~~~~~~~~~~~~~~~~~~~~~~~~~~& 
\irule{\Gamma \vdash B}
      {\Gamma \vdash C}
      {\begin{array}{l} \mbox{$\langle x,A,t \rangle$ $\fa$-elim}\\ \mbox{if $B \equiv (\fa x~A)$ and $C
         \equiv [t/x]A$}\end{array}}
\\
\\

\irule{\Gamma \vdash C}
        {\Gamma \vdash B}
        {\begin{array}{l} \mbox{$\langle x,A,t \rangle$ $\ex$-intro}\\ \mbox{if $B \equiv (\ex
        x~A)$ and $C \equiv [t/x]A$}\end{array}}
~~~~~~~~~~~~~~~~~~~~~~~~~~~~~~~~~~~~~~~~~~~~~~~~~~~~~& 
\irule{\Gamma \vdash C~~~\Gamma, A \vdash B}
        {\Gamma \vdash B}
        {\begin{array}{l} \mbox{$\langle x,A \rangle$ $\ex$-elim}\\ \mbox{ if $C \equiv (\ex x~A)$}\\ \mbox{and 
               $x \not\in FV(\Gamma B)$}\end{array}}
\end{array}$$
\caption{Natural Deduction Modulo Theory\label{ndm}}
\end{figure}

For instance, consider the congruence defined by the {\em subset reduction rule}
$$x \subseteq y \lra \fa z~(z \in x \Rightarrow z \in y)$$
The sequent $\vdash s \subseteq s$ has the proof
$$\irule{\irule{\irule{}
                      {z \in s \vdash z \in s}
                      {\mbox{axiom}}
               }
               {\vdash z \in s \Rightarrow z \in s}
               {\mbox{$\Rightarrow$-intro}}
        }
        {\vdash s \subseteq s}
        {\mbox{$\langle z, z \in s \Rightarrow z \in s\rangle$~$\fa$-intro}}$$

Note that when two propositions $A$ and $B$ are provably
equivalent, that is when $A \Leftrightarrow B$ is provable,
then the proposition $A$ has a proof if and only if the
proposition $B$ has a proof, but the propositions $A$ and $B$
need not have the same proofs.  In contrast, when two
propositions are congruent, that is when $A \equiv B$, then
every proof of $A$ is a proof of $B$ and vice versa, thus the
propositions $A$ and $B$ have the same proofs.

Sequent calculus modulo theory can be defined in the same way:
the rule
$$\irule{\Gamma \vdash A~~~\Gamma, B \vdash  \Delta}
        {\Gamma, A \Rightarrow B \vdash \Delta}
        {\mbox{$\Rightarrow$-left}}$$
for instance, is transformed into 
$$\irule{\Gamma \vdash A~~~\Gamma, B \vdash  \Delta}
        {\Gamma, C \vdash \Delta}
        {\begin{array}{l} \mbox{$\Rightarrow$-left}\\ \mbox{if $C \equiv (A \Rightarrow B)$}\end{array}}$$
where the proposition $A \Rightarrow B$ is replaced by any proposition $C$ 
such that $C \equiv (A \Rightarrow B)$. And the other rules are transformed
alike. See, for instance, \cite{DowekWerner} for a description of the full 
system.

Another variant of Natural deduction modulo theory and Sequent
calculus modulo theory is {\em Super-deduction}
\cite{Wack,KirchnerBraunerHoutmann}.  In Super-deduction, new
deduction rules are computed from the reduction rules.  For instance,
the subset reduction rule
yields the deduction rules 
$$\irule{\Gamma, z \in x \vdash z \in y}
        {\Gamma \vdash x \subseteq y}
        {\mbox{$z \not\in FV(\Gamma)$}}$$
$$\irule{\Gamma \vdash x \subseteq y~~~\Gamma \vdash z \in x}
        {\Gamma \vdash z \in y}
        {}$$
These rules are closer to the informal mathematical style than, for 
instance, Natural deduction rules. Indeed, to prove $x \subseteq y$, we often 
consider a generic element in $x$ and prove that it is in $y$ without 
using the universal quantifier and the implication of the proposition
$\fa z~(z \in x \Rightarrow z \in y)$. 
The fact that these derived rules use atomic propositions only also 
explains why connectives and quantifiers are less often 
used in informal proofs than in formal ones.

\subsection{Polarized deduction modulo theory}

In Natural deduction modulo theory and in Sequent calculus modulo
theory, the reduction rules are just used to define the congruence
$\equiv$. In fact, this congruence does not even need to be defined
with reduction rules and it could be any congruence, provided it is
decidable and it does not identify non-atomic propositions with
different head symbols.  But we may also want to stress that
computation is oriented and take, in these rules, the condition $C
\lra^* (A \Rightarrow B)$ instead of $C \equiv (A \Rightarrow B)$,
meaning that in the sequent $\Gamma, C \vdash \Delta$, the proposition $C$
can only be reduced.

In particular, the axiom rule 
$$\irule{}
        {\Gamma, A \vdash B}
        {\begin{array}{l} \mbox{axiom}\\ \mbox{if $A \equiv B$}\end{array}}$$
would be restated
$$\irule{}
        {\Gamma, A \vdash B}
        {\begin{array}{l} \mbox{axiom}\\ \mbox{if $A \lra^* C$ and $B \lra^* 
         C$}\end{array}}$$
If the theory contains rewrite rules on terms only, and 
$t$ and $u$ are two terms such that $t \equiv u$, it is still possible 
to prove the sequent $P(t) \vdash P(u)$. But when $t$ and $u$ do not 
have a common reduct, the proof of $P(t) \vdash P(u)$ contains cuts.
In other words, in this particular case, the Sequent calculus modulo theory 
has the cut elimination property if and only if the reduction system is
confluent \cite{confluence} and Newman's algorithm \cite{Newman}---which 
permits transforming
an equational proof into a valley proof---is a cut-elimination 
algorithm.

This idea of using a rewrite relation rather than a congruence in the
deduction rules can be developed further: the subset reduction rule
permits to prove the equivalence
$$x \subseteq y \Leftrightarrow \fa z~(z \in x \Rightarrow z \in y)$$
Thus, when the atomic proposition $P$ reduces to the proposition 
$A$, $P$ and $A$ must be equivalent.
For instance, it is not possible to reduce
$\mbox{\em Equilateral}(x)$ to $\mbox{\em Isosceles}(x)$ because a 
triangle may be isosceles without being equilateral.

More generally, it is easy to transform an axiom of the form 
$P \Leftrightarrow A$ into a reduction rule $P \lra A$, but, although 
it is possible \cite{BurelKirchner}, it is not 
easy to transform an axiom of the form 
$P \Rightarrow A$ into a reduction rule. 
We want to replace such an axiom with a rule that permits reducing $P$ into $A$
when $P$ is a hypothesis, but not when it is a goal.

This leads to an extension of Deduction modulo theory, called {\em
Polarized deduction modulo theory} where reduction rules are
classified into positive and negative, the positive rules may apply to
the positive occurrences of atomic propositions and the negative ones to
the negative occurrences. 

For instance, in Polarized sequent calculus modulo theory, the left rule of the 
implication is stated 
$$\irule{\Gamma \vdash A~~~\Gamma, B \vdash  \Delta}
        {\Gamma, C \vdash \Delta}
        {\begin{array}{l} \mbox{$\Rightarrow$-left}\\ \mbox{if $C \lra_-^*
(A \Rightarrow B)$}\end{array}}$$
and its right rule 
$$\irule{\Gamma, A \vdash B}
        {\Gamma \vdash C}
        {\begin{array}{l} \mbox{$\Rightarrow$-right}\\ \mbox{if $C \lra_+^*
(A \Rightarrow B)$}\end{array}}$$

Polarized deduction modulo theory is the flavor of Deduction modulo theory 
that is more often used in automated theorem proving. 

The first reason is 
that, in clause based theorem proving, a reduction rule of the form 
$$x \in y \cup z \lra x \in y \vee x \in z$$
can be used to reduce a positive literal in a clause 
but not a negative one.
For instance, the clause $L_1 \vee L_2 \vee a \in b \cup c$ reduces to 
the clause $L_1 \vee L_2 \vee a \in b \vee a \in c$, 
but the clause $L_1 \vee L_2 \vee \neg a \in b \cup c$ reduces to 
the proposition 
$L_1 \vee L_2 \vee \neg (a \in b \vee a \in c)$ that is not a clause. 
In contrast, if we replace this reduction rule by the polarized rules
$$x \in y \cup z \lra_- x \in y \vee x \in z$$
$$x \in y \cup z \lra_+ x \in y$$
$$x \in y \cup z \lra_+ x \in z$$
then the clause $L_1 \vee L_2 \vee \neg a \in b \cup c$ reduces to the clauses
$L_1 \vee L_2 \vee \neg a \in b$ and to 
$L_1 \vee L_2 \vee \neg a \in c$.
More generally, any reduction system can be transformed this way
to a clausal one \cite{Gao}.

The second reason is that any consistent set of axioms can be transformed 
into a Polarized reduction system that is classically equivalent 
\cite{stacs,Burel} and some sets of axioms can be transformed into 
a Polarized reduction system that is constructively equivalent 
\cite{Burel9}.

Interestingly, this result has been proved with applications
to automated theorem proving in mind, it uses automated theorem proving 
methods, but it is a purely proof-theoretical result.

\subsection{Expressing theories in Deduction modulo theory}

The early work on expressing theories in Deduction modulo theory 
was focused on specific theories: Simple type theory \cite{DHKHOL}, 
Arithmetic \cite{Peano,Allali}, Set theory \cite{DowekMiquel}, etc.

Then, as already said, systematic ways of transforming sets of axioms
into sets of reduction rules have been investigated
\cite{stacs,Burel,Burel9}.

\subsection{The $\lambda \Pi$-calculus modulo theory}

The early developments of Deduction modulo theory were 
independent of the proofs-as-algorithms paradigm, also 
known as the Brouwer-Heyting-Kolmogorov interpretation, that is the idea 
that a proof of $A \Rightarrow B$, for instance, is an algorithm transforming 
proofs of $A$ into proofs of $B$.
In Deduction modulo
theory, like in Predicate logic, 
terms, propositions, and proofs belong to three different
languages, and proofs are not terms. But we have mentioned that one
of the origins of Deduction modulo theory was the definitional equality of
Martin-L\"of's Intuitionistic type theory. This suggests that this
idea of identifying congruent propositions can also be useful in
systems based on the proofs-as-algorithms paradigm.

The simplest system to express proofs of Predicate logic as algorithms is the
$\lambda$-calculus with dependent types \cite{HHP}, also know as the $\lambda
\Pi$-calculus. This leads to the development of an extension of the
$\lambda \Pi$-calculus, called the $\lambda \Pi$-calculus modulo
theory \cite{CousineauDowek}. This system is closely related to 
Martin-L\"of's logical framework \cite{NPS}. 

Any theory that can be expressed in minimal Deduction modulo theory,
that is in the restriction of Deduction modulo theory, where the only
logical operators are the implication and the universal quantifier,
can be expressed in the $\lambda \Pi$-calculus modulo theory. In
particular Simple type theory can be expressed in the $\lambda
\Pi$-calculus modulo theory. An interesting point here is that the
Calculus of Constructions \cite{CoquandHuet} has been designed to
express proofs of Simple type theory as algorithms.  It happens that
$\lambda \Pi$-calculus modulo theory also can express those proofs as
algorithms. This suggests that the Calculus of Constructions itself
could be expressed in the $\lambda \Pi$-calculus modulo theory, and
this is indeed the case \cite{CousineauDowek}. The embedding of the
Calculus of Constructions into the $\lambda \Pi$-calculus modulo
theory follows closely the expression of Simple type theory in
Deduction modulo theory.

It happens {\em a posteriori} that this embedding of the 
Calculus of Constructions into the $\lambda \Pi$-calculus modulo theory 
can be seen as an extension of the $\lambda \Pi$-calculus with an 
impredicative universe {\em \`a la} Tarski \cite{Assaf} 
and thus that there is a strong 
link between the expression of Simple type theory in Predicate logic and 
the notion of universe {\em \`a la} Tarski.

\section{Properties}

\subsection{Models}

The usual models of classical Predicate logic, valued in $\{0,1\}$, 
can be used for Deduction modulo theory. A congruence $\equiv$ is said 
to be valid in a model when $A \equiv B$ implies 
$\llbracket A \rrbracket_{\phi} =
\llbracket B \rrbracket_{\phi}$ for all valuations $\phi$, and a
soundness and completeness theorem can be proved using standard 
methods.

Like for Predicate logic, the set of truth values $\{0,1\}$ can be
extended to any Boolean algebra, allowing to prove a stronger
completeness theorem: given a theory, 
there exists a model such that the propositions
valid in this model are exactly the propositions provable in this
theory.

Boolean algebras can be extended to Heyting algebras to define 
a sound and complete semantics for constructive logic.

However, in all these models---valued in $\{0,1\}$, in Boolean algebras and 
in Heyting algebras---, two provably equivalent propositions always have the
same truth value: if $A \Leftrightarrow B$ is valid, then $A
\Rightarrow B$ and $B \Rightarrow A$ are valid, hence $\llbracket A
\rrbracket_{\phi} \leq \llbracket B \rrbracket_{\phi}$ and $\llbracket
B \rrbracket_{\phi} \leq \llbracket A \rrbracket_{\phi}$ and by
antisymmetry $\llbracket A \rrbracket_{\phi} = \llbracket B
\rrbracket_{\phi}$.  Thus, there is no way to make a difference, in the
model, between provable equivalence and congruence: whether $A$ and $B$
are just equiprovable or have the same proofs, they have the same
truth value.

A way to overcome this is to extend Boolean algebras and Heyting
algebras by dropping the antisymmetry condition on the relation
$\leq$. This relation is then a pre-order and the algebras defined
this way can be called {\em pre-Boolean} algebras \cite{BHH} and {\em
  pre-Heying} algebras \cite{TVA}. The soundness theorem is
proved exactly the same way---antisymmetry is never used in this
proof---, and the completeness is simpler as the class of models is
larger. This corresponds to the intuition that the relation
$\leq$, defined by $A \leq B$ if $A \Rightarrow B$ is provable,
is reflexive and transitive, but not antisymmetric.

This way, two provably equivalent propositions may be interpreted by 
distinct truth values, unlike two congruent propositions that must 
be interpreted by the same one, and it is possible to define models 
where a proposition $A$ is interpreted by the set of its proofs. 

When a theory has a model valued in some pre-Heyting algebra 
it is consistent, when it has a model valued in all pre-Heyting algebras
it is said to be {\em super-consistent}. 

\subsection{Cut-elimination}

Proof-reduction is defined in Deduction modulo theory in the same way as 
in Predicate logic, but the difference is that it does not 
always terminate.
Indeed, if we define a theory with the reduction rule 
$P \lra (P \Rightarrow Q)$ 
the sequent $\vdash Q$ has the following proof 
$$\irule{\irule{\irule{\irule{}
                             {P \vdash P \Rightarrow Q}
                             {\mbox{axiom}}
                       ~~~~~~~~~~
                       \irule{}
                             {P \vdash P}
                             {\mbox{axiom}}
                      }
                      {P \vdash Q}
                      {\mbox{$\Rightarrow$-elim}}
               }
               {\vdash P \Rightarrow Q}
               {\mbox{$\Rightarrow$-intro}}
         ~~~~~~~~~~~~~~~~~~~~~~~~~~~~~~~~~~~~~~~
         \irule{\irule{\irule{}
                             {P \vdash P \Rightarrow Q}
                             {\mbox{axiom}}
                       ~~~~~~~~~~
                       \irule{}
                             {P \vdash P}
                             {\mbox{axiom}}
                      }
                      {P \vdash Q}
                      {\mbox{$\Rightarrow$-elim}}
               }
               {\vdash P}
               {\mbox{$\Rightarrow$-intro}}
        }
        {\vdash Q}
        {\mbox{$\Rightarrow$-elim}}$$
that contains a cut and that reduces to itself.

Moreover, it is possible to prove that all cut free, that is
irreducible, proofs end with an introduction rule, thus not only this
proof does not terminate, but the sequent $\vdash Q$ has no cut free
proof.

And a similar example can be built with a terminating
reduction system \cite{DowekWerner}.

Not only some theories have the cut-elimination property and some others
do not, but this property is even undecidable 
\cite{BurelKirchner,Hermantperso}. 

Thus, unlike for axiomatic theories, the notion of proof-reduction can be 
defined in a generic, theory independent, way, and the properties of 
cut free proofs, such as the property that the last rule of a cut free
proof is an introduction rule can be proved in a generic way. But, the 
proof-termination theorem itself must be proved for each theory.

Using a method introduced to prove the termination of proof reduction
in Simple type theory \cite{Girard}, 
we can prove that proof-reduction terminates in 
some theory, if a reducibility candidate $\llbracket A \rrbracket$
can be associated to each proposition $A$, 
in such a way that two congruent propositions 
are associated with the same reducibility candidate \cite{DowekWerner}
$$A \equiv B~\mbox{implies}~\llbracket A \rrbracket = \llbracket B \rrbracket$$

This association of a reducibility candidate to each proposition is
thus a model valued in the algebra of the reducibility candidates and
the condition that two congruent propositions are associated with the
same reducibility candidate is the validity of this congruence in this
model.

This way, we get that if a theory has a model valued in the algebra of
reducibility candidates, then proof-reduction strongly terminate.

The algebra of reducibility candidates is a pre-Heyting algebra---but
not a Heyting algebra---thus we also get that proof-reduction
terminates in super-consistent theories.

This semantic view on termination of proof reduction theorems also
permits to relate these termination proofs to the so called {\em
  semantic} cut-elimination proofs that proceed by proving a
completeness result for cut free provability.  First, without proving
the termination of proof-reduction, it is possible to prove directly
that, in a super-consistent theory, each provable proposition has a
cut free proof \cite{DowekHermant,BHH}.  This completeness proof does not
use the pre-Heyting algebra of reducibility candidates but a simpler
one.

Then, in some non super-consistent theories, proof reduction does not
terminate, but each provable proposition has a cut free proof
\cite{Hermant05}. An example is obtained by replacing the proposition
$Q$ by $\top$ in the example above.  This proof still fails to
terminate but the sequent $\vdash \top$ has another proof, that is cut
free. Such cut-elimination theorems can only be proved via a
completeness theorem and, when they are proved constructively, the
constructive content of these proofs is a proof-transformation
algorithm, that need not be related to proof-reduction.

Finally, some theories do not have the cut elimination property, but
they can sometimes be extended to theories that have this property by adding
derivable reduction rules \cite{BurelKirchner,Burel14}. 
This saturation process can be
compared to Knuth-Bendix method \cite{KB}---remember that confluence
is a special case of cut-elimination---that does not prove that all reduction
systems are confluent, but that, in some cases, it is possible to
extend a reduction system with derivable rules, to make it confluent.

\subsection{Automated theorem proving methods}

Deduction modulo theory has been introduced to design and study
automated theorem proving methods. The first method introduced was a
variant of Resolution \cite{DHK} that was too complicated because rules
were not polarized. Thus, clauses could rewrite to non clausal
propositions that needed to be dynamically transformed into clausal form.
Polarization permitted to
simplify the method \cite{polar} and also to understand
better its relation to other methods. This method is complete if and only
if the theory has the cut-elimination property \cite{Hermant10}.

Imagine we have a clause $$L_1 \vee L_2 \vee a \in b \cup c$$
and a negative reduction rule 
$$x \in y \cup z \lra_- x \in y \vee x \in z$$
then applying this rule to this clause yields the clause
$$L_1 \vee L_2 \vee a \in b \vee a \in c$$
But instead of this reduction rule, we could have taken a clause 
$$\underline{\neg x \in y \cup z} \vee x \in y \vee x \in z$$ and
Resolution, applied to the literal $a \in b \cup c$ and the underlined
literal in the new clause, would have yielded the same result. Thus,
there is no need to extend Resolution to handle reduction rules, but
reduction rules can just be seen as special clauses, called {\em
  one-way clauses}. The Resolution rule cannot be applied to two
one-way clauses and when it is applied to a one-way clause and an
ordinary one, only the literal corresponding to the left-hand side of
the reduction rule can be used. Thus, Polarized resolution modulo
theory is just another variant of Equational resolution with clause
restrictions---like the Set of support \cite{Wos} and the Semantic
resolution \cite{Slagle} strategies---and literal restrictions---like
Ordered resolution.

But, unlike other variants of Resolution, its completeness is
equivalent to a cut-elimination theorem. Thus, it permits to handle
theories, such as Simple type theory, that cannot be handled, for
instance, with Ordered resolution, as the completeness of Polarized
resolution modulo the rules of Simple type theory implies cut
elimination for Simple type theory and, unlike that of Ordered
resolution, it cannot be proved in Simple type theory
\cite{BurelDowek}.

A side effect of this work is to show that, surprisingly, clause
restriction strategies---such as the Set of support or the Semantic
resolution strategy---and literal restriction strategies---such as
Ordered resolution---can be combined, provided we do not consider
theories that are just consistent, but theories that also have the cut
elimination property.

These remarks also showed the way to combine this method with other
selection strategies in Resolution. In particular, it has been shown
that this restriction is compatible with Ordered resolution
\cite{Burel10}.

Besides Resolution, other proof-search methods have been investigated, 
in particular direct search in cut free sequent calculus modulo theory, 
also known as the {\em tableaux} method \cite{BonichonHermant}.

\section{Implementations}

The early work on Deduction modulo theory only led to experimental 
implementations. But more mature systems have been developed in the 
recent years.

\subsection{Dedukti}

Dedukti \cite{Boespflug,BCH,Saillard} 
is an implementation of the $\lambda\Pi$-calculus modulo
theory.  It is thus based on the proofs-as-algorithms paradigm and
proof-checking is reduced to type-checking. But type-checking itself
may require an arbitrary amount of computation to check the congruence
of two propositions.

Dedukti is a parametric system: by changing the reduction rules, we
change the theory in which the proofs are checked. Thus Dedukti is a
logical framework \cite{HHP}. As the proofs of many different systems
can be expressed in this framework Dedukti is mostly used to check
proofs developed in other systems---hence its name: ``to deduce'' in
Esperanto---: HOL \cite{Assafmaster}, Focalize \cite{Cauderlier}, Coq
\cite{BoespflugBurel,Assaf}, etc. as well as proofs produced by
automated theorem proving systems, such as iProver, Zenon, iProver
modulo, and Zenon modulo.  The current goal of the project is to be
able to interface proofs developed in different systems, and defining
a standard for proofs in various theories, much the same way standards
are defined, for instance, for SMT solvers
\cite{BessonFontaineThery,StumpOeReynoldsHarareanTinelli}.

\subsection{iProver modulo, Super Zenon and Zenon modulo}

iProver modulo \cite{Burel11} 
is an implementation of Ordered polarized resolution modulo theory.
It is developed as an extension of iProver.
It has shown convincing experimental results compared to the axiomatic
approach. A tool automatically orienting axioms into rewriting systems
usable by iProver Modulo is also available.

Super Zenon \cite{JBDD} 
is an implementation of Tableaux modulo theory specifically 
designed for a variant of Class theory---Second order logic---called 
{\em B set theory}, and using Super-deduction instead of the original 
Deduction modulo theory.

Zenon modulo \cite{DDGHH1,DDGHH}
is a generic implementation of the Tableaux modulo theory method. 
It comes with a heuristic that turns axioms into rewrite rules before 
performing proof-search, and also with a new hand-tailored expression of 
B set theory as a set of rewrite rules.

\section{Trends and Open Problems}

In recent years, the effort in Deduction modulo theory has been put on the
development of implementations. In particular, we do not know how far we can 
go in interfacing proof systems using a logical framework such as Dedukti.
We also need to investigate how having user defined reduction rules can 
impact tactic based proof development.

In automated theorem proving we do not understand yet how to mix Resolution
modulo theory with equality specific methods such as superposition.

On the more proof-theoretical side, we know that super-consistency is
a sufficient condition for the strong termination of proof reduction
but we do not know if it is a necessary condition. As suggested in
\cite{Cousineau}, the notion of super-consistency may require some
adjustment so that we can prove that it is a necessary and sufficient
condition for proof termination.
Finally, some extension of Deduction modulo theory
allow congruences that identify non-atomic propositions with different 
head-symbols \cite{DD}, in particular isomorphic types such as 
$A \Rightarrow (B \wedge C)$ and $(A \Rightarrow B) \wedge (A \Rightarrow C)$,
but we do not know yet how far we can go in this direction.

\bibliographystyle{plain}
\bibliography{Bibliography}

\end{document}